\documentclass[letterpaper, 10 pt, conference]{ieeeconf}  

\IEEEoverridecommandlockouts                              
\usepackage{amsmath,amssymb,physics,mathtools} 
\usepackage{xcolor}
\usepackage{pgfplots}
\usepackage{pgfplotstable}
\usepgfplotslibrary{fillbetween}
\usepackage[capitalize]{cleveref}
\let\labelindent\relax
\usepackage{enumitem}

\usepackage{amsthm}
\usepackage{algorithmicx}
\usepackage{algorithm}
\usepackage{algpseudocode}
\usepackage{placeins} 

\usepackage{url}
\usepackage{mdframed}
\usepackage{cite}
\makeatletter
\def\@cite#1#2{[{#1\if@tempswa , #2\fi}]}

\makeatother
\usepackage{subcaption}
\usepackage{wrapfig}
\usepackage[tracking=false,kerning=true,spacing=true]{microtype}

\usepackage{siunitx}
\DeclareSIUnit{\watthour}{Wh}

\usepackage{tikz,pgfplots}
\usetikzlibrary{arrows.meta, positioning, calc}
\pgfplotsset{every axis/.append style={
    tick label style={font=\footnotesize},
    legend style={font=\footnotesize},
    label style={font=\small}
}}

\definecolor{ETHblue}{RGB}{0,112,192}      
\definecolor{ETHpurple}{RGB}{128,0,128}      
\definecolor{ETHpetrol}{RGB}{0,153,153}      
\definecolor{ETHgreen}{RGB}{112,173,71}      
\definecolor{ETHbronze}{RGB}{112,79,18}      
\definecolor{ETHrot}{RGB}{150,39,45}         
\definecolor{ETHpurpur}{RGB}{140,10,89}       
\definecolor{ETHgrau}{RGB}{87,87,87}          

\newtheoremstyle{DefinitionStyle}
{5pt}{5pt}                
{}                        
{}                        
{\bfseries}               
{}                       
{3pt plus 1pt minus 1pt}  
{%
  \thmname{#1}\thmnumber{ #2 }\thmnote{\normalfont(#3)}
  \;
}
\theoremstyle{DefinitionStyle}
\newtheorem{definition}{Definition}

\newtheoremstyle{AssumptionStyle}
  {5pt}{5pt}                
  {}                        
  {}                        
  {\bfseries}               
  {}                        
  {3pt plus 1pt minus 1pt}  
  {%
    \thmname{#1}\thmnumber{ #2 }\thmnote{\normalfont(#3)}
    \;
  }

\theoremstyle{AssumptionStyle}
\newtheorem{assumption}{Assumption}
\crefname{assumption}{Assumption}{Assumptions}

\newtheoremstyle{Theorem}
  {5pt}{5pt}                
  {}                        
  {}                        
  {\bfseries}               
  {}                        
  {3pt plus 1pt minus 1pt}  
  {%
    \thmname{#1}\thmnumber{ #2 }\thmnote{\normalfont(#3)}
    \;
  }

\theoremstyle{TheoremStyle}
\newtheorem{theorem}{Theorem}
\newtheoremstyle{PropositionStyle}
  {5pt}{5pt}                
  {}                        
  {}                        
  {\bfseries}               
  {}                        
  {3pt plus 1pt minus 1pt}  
  {%
    \thmname{#1}\thmnumber{ #2 }\thmnote{\normalfont(#3)}
    \;
  }

\newtheorem{proposition}{Proposition}
\newtheoremstyle{CorollaryStyle}
  {5pt}{5pt}                
  {}                        
  {}                        
  {\bfseries}               
  {}                        
  {3pt plus 1pt minus 1pt}  
  {%
    \thmname{#1}\thmnumber{ #2 }\thmnote{\normalfont(#3)}
    \;
  }

\newtheorem{corollary}{Corollary}
\newtheoremstyle{Examplestyle}
  {5pt}{5pt}                
  {\normalfont}             
  {}                        
  {\bfseries}               
  {}                        
  {3pt plus 1pt minus 1pt}  
  {%
    \thmname{#1}\thmnumber{ #2 }\thmnote{\normalfont(#3)}
    \;
  }

\theoremstyle{Examplestyle}
\newtheorem{example}{Example}
\DeclareMathOperator*{\argmin}{argmin}

\newcommand{\reals}{\mathbb{R}}
\newcommand{\nonnegativereals}{\mathbb{R}_{\geq 0}}
\newcommand{\positivereals}{\mathbb{R}_{>0}}

\newcommand{\aggregatedistribution}{\mu}
\newcommand{\aggregatevar}{\hat\sigma}
\newcommand{\aggregate}{\sigma}
\newcommand{\aggregatekth}{\sigma_k}

\newcommand{\game}{$\mathcal{G}(\{\mathcal{X}^i,J^i\}_{i=1}^N,\sigma)$}

\newcommand{\pnorm}{p}
\renewcommand{\d}{\mathrm{d}}
\newcommand{\wardropeq}{x_W}
\newcommand{\aggregatespace}{\mathcal{X}_\sigma}

\newcommand{\sav}[1]{\textcolor{red}{[#1]\raise 0.5ex \hbox{\footnotesize{SB}}}}

\newcommand{\revision}[1]{\textcolor{black}{#1}}

\title{\LARGE \bf
\revision{Strategically Robust Aggregative Games}
}

\author{Andreas Feik, Nicolas Lanzetti, Saverio Bolognani, Florian Dörfler, and Dario Paccagnan 
\thanks{A. Feik, S. Bolognani, and F. Dörfler are with the Automatic Control Laboratory, ETH Zurich, 8092 Zurich, Switzerland,
        {\tt\small \{anfeik,bsaverio,dorfler\}@ethz.ch}}%
\thanks{N. Lanzetti is with the Department of Computing and Mathematical Sciences, Caltech, Pasadena, USA,
        {\tt\small lnicolas@caltech.edu}}%
\thanks{D. Paccagnan is with the Department of Computing, Imperial College London, London, UK,
        {\tt\small d.paccagnan@imperial.ac.uk}}%
\thanks{This work was supported by the NCCR Automation, a National Centre of Competence in Research, funded by the Swiss National Science Foundation (grant number 51NF40\textunderscore225155).}
}

\begin{document}

\maketitle
\thispagestyle{empty}
\pagestyle{empty}

\begin{abstract}
In many multiagent settings, such as electric vehicle charging and traffic routing, agents must make decisions in the face of uncertain behavior exhibited by others. Often, this uncertainty arises from multiple sources, such as incomplete information, limited computation, or bounded rationality, ultimately impacting the aggregate behavior.
%
To tackle this challenge, we follow recent work on \emph{strategically robust} game theory and \revision{postulate} that agents seek protection directly against deviations around the emergent behavior, as opposed to explicitly modeling all sources of uncertainty.
Specifically, we propose that each agent protects itself against the worst-case aggregate behavior within an optimal-transport-based ambiguity set centered at the emergent aggregate population behavior.
This leads to a novel equilibrium concept, called strategically robust Wardrop equilibrium,
%
that enables \revision{one} to interpolate between standard Wardrop equilibria (no robustness) \revision{and} security strategies (maximum robustness). 
In the setting of convex aggregative games, we establish the existence of a pure strategically robust Wardrop equilibrium and provide tractable computational tools for computing it. 
Through an application in electric vehicle charging, we demonstrate that strategically robust Wardrop equilibria lead to better decisions, protecting agents against the uncertain aggregate behavior of the population. 
Remarkably, we also observe that strategic robustness can lead to lower equilibrium costs for all agents, uncovering a ``coordination-via-robustification'' effect.

\end{abstract}


\section{Introduction}\label{sec:introduction}

In many of today's large-scale systems, agents must make decisions while uncertain about others' aggregate behavior.
For instance, in a traffic network, agents must route themselves to their corresponding destinations. However, perfect anticipation of the expected aggregate behavior, which impacts travel times and costs, is often impossible.
It is therefore natural for agents to seek protection against this uncertainty. 
When this uncertainty is \emph{exogenous} \revision{(e.g., an unknown disturbance)}, agents can model (or estimate) it and incorporate it into their decision-making. 
For this, various approaches have been proposed, \revision{such as robust games and, more recently, distributionally robust games;}
e.g., see \cite{aghassi2006robust,ordonez2010wardrop,loizou2015distributionally,bauso2017distributionally,peng2021games,pantazis2024nash,pantazis2023data,wood2024solving}.

\revision{However, this uncertainty is often \emph{endogenous}. Agents are uncertain about the aggregate behavior of the population, which they themselves impact, both directly (because the aggregate behavior includes each agent's behavior) and indirectly (because each agent's behavior impacts that of others).}
Additionally, the nature of the uncertainty is rarely known; 
e.g., uncertain behavior may emerge due to bounded rationality, imperfect knowledge, and limited computation. Given the presence of all these sources of uncertainty, it is natural for the agents to seek decisions that protect themselves against deviations around the aggregate behavior itself rather than trying to explicitly model each of them.

\medskip

In this paper, we follow this approach initiated in~\cite{lanzetti2025strategically}, and propose that each agent takes decisions in the face of the worst-case aggregate behavior within a suitably-defined ambiguity set, which is itself centered around the emergent aggregate behavior that is so implicitly defined.
Motivated by its expressivity and computational properties, we construct ambiguity sets via optimal transport. These ambiguity sets contain all aggregate behaviors (i.e., probability distributions over \revision{possible aggregates}) that are not too dissimilar from the emergent aggregate behavior, as measured through the so-called \emph{Wasserstein distance}~\cite{villani2008optimal}.
The radius of this ambiguity set allows us to tune the level of robustness. When the radius collapses to $0$, the ambiguity set only contains the original aggregate behavior, and we recover the classical notion of Wardrop equilibrium. When, instead, the radius approaches infinity, players protect themselves against \emph{any} aggregate behavior, thereby recovering the \mbox{notion of security strategy.}

\medskip

We present our results in the setting of $N$-player convex aggregative games, widely used for interactions in large-scale systems with strategic agents, e.g., see~\cite{paccagnan2018nash} and references therein.
This setting effectively allows us to expand the scope of ~\cite{lanzetti2025strategically}, which instead focuses on Nash equilibrium problems, in three ways. 
First, strategic robustness as proposed in~\cite{lanzetti2025strategically} requires knowledge about each agent's action, which is impractical in large-scale systems. Thanks to the aggregate setting, we instead require knowledge about the aggregate behavior. 
Second, the computational complexity of~\cite{lanzetti2025strategically} scales with the number of agents. This makes the analysis of games with many agents challenging. The aggregative formulation we propose, instead, allows us to design algorithms whose computational cost is independent of the number of agents. 
Third, robustifying against the aggregate can capture effects such as that of a new player entering the game, which cannot be easily modeled in a standard $N$-player game.


\medskip

\paragraph*{Contributions}
Our contributions are threefold. 

\begin{mdframed}[hidealllines=true,backgroundcolor=blue!10]
\textbf{C1.} We introduce the notion of strategically robust Wardrop equilibria and, for convex games, we prove existence of a \emph{pure} strategically robust Wardrop equilibrium (\cref{sec:strategically_robust_wardrop_equilibrium:definition,sec:strategically_robust_wardrop_equilibrium:existence}).
\end{mdframed}
Existence of standard (pure) Wardrop equilibria usually follows from tools from variational inequalities~\cite{facchinei2003finite}, which however demand differentiability of the costs.
Unfortunately, worst-case costs
generally fail to meet this property.
%
Thus, we resort to Kakutani's fixed-point theorem, which solely relies on continuity arguments, together with Berge's Maximum Theorem and the composition properties of set-valued maps. 

\begin{mdframed}[hidealllines=true,backgroundcolor=blue!10]
\textbf{C2.} 
We show that the problem of finding a strategically robust Wardrop equilibrium can be reformulated into the problem of finding \emph{standard} Wardrop equilibria of an appropriately modified convex aggregative game, which is finite-dimensional
(\cref{sec:strategically_robust_wardrop_equilibrium:equivalent_game,sec:strategically_robust_wardrop_equilibrium:solution}).
%

%
\end{mdframed}

The computation of strategically robust Wardrop equilibria appears challenging; e.g.,
the mere evaluation of worst-case costs is an infinite-dimensional optimization problem over the space of probability distributions. 
\revision{We bypass this complexity as in strategically robust game theory~\cite[Section 4]{lanzetti2025strategically}: 
strategically robust Wardrop equilibria are standard Wardrop equilibria of an appropriately modified convex aggregative game.
While the cost function of this modified game involves solving an optimization problem, we use duality in distributionally robust optimization to further reformulate it as a standard convex aggregative game with augmented action spaces. This way, we can compute strategically robust Wardrop equilibria using standard equilibrium-seeking algorithms.} 

%
%

\begin{mdframed}[hidealllines=true,backgroundcolor=blue!10]
\textbf{C3.} We study the role of strategic robustness in an electric vehicle charging game (\cref{sec:ev_charging}).
%
Our results show not only that strategic robustness protects agents against uncertainty, but also uncover a remarkable \emph{``coordination-via-robustification effect''} whereby strategic robustness yields lower equilibrium costs for \emph{all} agents.
In some settings, we even demonstrate that, when appropriately calibrated, strategic robustness can coordinate agents to the social optimum, driving the price of anarchy to 1. 
\end{mdframed}

\section{Strategically Robust Wardrop Equilibria}

\revision{We consider a game in which each player $i\in\{1,\ldots,N\}$ takes action $x^i\in\mathcal{X}^i\subset \mathbb{R}^n$. Each agent aims to minimize the cost $J^i(x^i,\aggregate)$, which depends on their action $x^i$ and the aggregate $\aggregate:\mathcal{X}^1\times\cdots\times\mathcal{X}^N\to\mathbb{R}^n$ of all players' actions, e.g., the ``average'' $\aggregate=\frac{1}{N}\sum_{i=1}^Nx^i$.
Since the cost only depends on an aggregate of all players' actions and not on the action of each player individually, the game is called \emph{aggregative}. We operate under the standard convex game assumptions:}

\begin{assumption}[Convex game]\label{assumption}
For each $i\in\{1,\ldots,N\}$: 
\begin{enumerate}
    \item The action space $\mathcal{X}^i$ is a non-empty compact and convex subset of $\mathbb{R}^n$.

    \item The function $x^i\mapsto J^i(x^i,\aggregate)$ is convex for all $\aggregate\in\mathbb{R}^n$. 

    \item The aggregate function $\aggregate:\mathcal{X}^1\times\cdots\times\mathcal{X}^N\to \mathbb{R}^n$ is continuous. 

    \item The cost $(x^i,\sigma)\mapsto J^i(x^i,\aggregate)$ is continuous. 
\end{enumerate}
\end{assumption}

We denote this game by \game.

\subsection{Strategically Robust Wardrop Equilibrium}\label{sec:strategically_robust_wardrop_equilibrium:definition}

We consider players that seek protection against uncertainty in the aggregate $\aggregate$. We suppose players know that the aggregate $\aggregate$ takes values in a non-empty set $\aggregatespace\subset\mathbb{R}^n$.

\begin{assumption}[Compactness]\label{assumption_compact}
    The set $\mathcal X_\aggregate$ is compact and satisfies $\aggregate(\mathcal{X}^1 \times \cdots \times \mathcal{X}^N)\subseteq \aggregatespace$.
\end{assumption}

When each player is aware of the others' action spaces, a natural choice is $\aggregatespace=\aggregate(\mathcal{X}^1 \times \cdots \times \mathcal{X}^N)$, which is compact being the image of a compact set under a continuous map.
\revision{In practice, however, players might not know the others' action spaces, so we present our results for a general compact set $\aggregatespace$ ensuring that $\aggregate(\mathcal{X}^1 \times \cdots \times \mathcal{X}^N)\subseteq \aggregatespace$.}

\subsubsection*{Ambiguity set}
To robustify against deviations in the aggregate behavior $\sigma(x)$, each player constructs an ambiguity set of probability distributions over the aggregate behavior, centered at the nominal one (i.e., the Dirac delta $\delta_{\sigma(x)}$).
We construct this ambiguity set via optimal transport: For $\pnorm\geq 1$, the type-$p$ Wasserstein distance between $\mu$ and $\delta_{\aggregate}$ is
\begin{align*}
    W_p(\mu,\delta_{\aggregate} )
    &=
    \min_{\gamma \in \Gamma(\mu, \delta_{\aggregate})}
    \left( \int_{\aggregatespace} \norm{\aggregate_1-\aggregate_2}^p \d\gamma(\aggregate_1, \aggregate_2) \right)^{\frac{1}{p}},
\end{align*}
where $\Gamma(\mu, \delta_{\aggregate})$ is the set of joint probability distributions whose first marginal is $\mu$ and whose second marginal is $\delta_{\aggregate}$. Intuitively, the Wasserstein distance quantifies the minimum cost to morph $\mu$ into $\delta_{\aggregate}$ when transporting a unit of mass from $\aggregate_1$ to $\aggregate_2$ costs $\norm{\aggregate_1-\aggregate_2}^p$, where $\norm{\cdot}$ is any norm \mbox{on $\aggregatespace$.}

The ambiguity set is defined as the set of all distributions whose distance from $\delta_{\aggregate}$, as measured through $W_p$, is at most~$\varepsilon$:
\begin{equation*}
    \mathcal{B}_{\varepsilon}(\delta_{\aggregate})
    \coloneqq 
    \{ \mu \in \mathcal{P}(\mathcal{X}_{\aggregate})
    :
    W_{\pnorm}(\mu, \delta_{\aggregate}) \leq \varepsilon \}.
\end{equation*}
The radius $\varepsilon$ allows us to tune the size of the ambiguity set and, therefore, the resulting level of robustness.
When $\varepsilon=0$, the ambiguity set only contains $\delta_{\aggregate}$; in this case, players are not robust. Conversely, when $\varepsilon\to+\infty$, the ambiguity set includes all possible distributions on $\aggregatespace$, meaning that players are protected against any possible aggregate behavior.

\subsubsection*{Equilibrium definition}
We now introduce the notion of strategically robust Wardrop equilibrium. In such an equilibrium allocation, each player selects an action minimizing the worst-case cost in the face of all distributions \mbox{in this ambiguity set.}

\begin{definition}[Pure strategically robust Wardrop equilibrium]\label{def:sr}
Given an aggregative game \game{} and a set $\aggregatespace$, 
the action profile $\wardropeq=(\wardropeq^1, \ldots, \wardropeq^N) \in \mathcal{X}^1\times\cdots\times\mathcal{X}^N$ is a \emph{pure strategically robust Wardrop equilibrium} with robustness level $\varepsilon\in\nonnegativereals$ if for all $i \in \{1, \ldots, N\}$ 
\begin{equation}\label{eq:defintion_sr_2}
    \wardropeq^i
    \in
    \argmin_{x^i\in \mathcal{X}^i} \sup_{\aggregatedistribution \in \mathcal{B}_{\varepsilon} (\delta_{\aggregate(\wardropeq)})} \mathbb{E}^{\aggregatevar \sim \aggregatedistribution} [J^i(x^i, \aggregatevar) ].
\end{equation}
In particular, when $\varepsilon=0$, we call $\wardropeq$ a Wardrop equilibrium.
\end{definition}

A few remarks on~\cref{def:sr} are in order. 

\paragraph*{Protection against mixed and pure aggregate behaviors} 
\revision{Our ambiguity sets contain \emph{probability distributions} over possible aggregate behaviors, and not only deterministic ones. This way, players are not only protected against different aggregate behaviors but also when the aggregate behavior is stochastic.}

\paragraph*{Interpolation between Wardrop and security strategies}
When $\varepsilon=0$, the ambiguity set $\mathcal{B}_{\varepsilon} (\delta_{\aggregate(\wardropeq)})$ collapses to $\{\delta_{\aggregate(\wardropeq)}\}$ and we recover the standard definition of Wardrop equilibrium. 
When instead $\varepsilon\to\infty$, the ambiguity set $\mathcal{B}_{\varepsilon} (\delta_{\aggregate(\wardropeq)})$ contains all possible aggregate behaviors and we recover security strategies. Thus, the radius of the ambiguity set interpolates between these two concepts. 

\paragraph*{Wardrop vs Nash}
The key distinction between our strategically robust Wardrop equilibrium and strategically robust equilibrium in \cite{lanzetti2025strategically} 
lies in the assumption that the effect of player $i$'s action $x^i$ on the aggregate $\aggregate(x)$ is negligible (e.g., because of a large population of agents), as in the standard setting of Wardrop equilibria. That is, when an agent deviates from $x^i$ to $\tilde x^i$, the aggregate action $\aggregate(x)$ does \emph{not} change and continues to define the center of our ambiguity sets. 

\paragraph*{On the equilibrium condition~\eqref{eq:defintion_sr_2}}
The strategically robust Wardrop equilibrium condition~\eqref{eq:defintion_sr_2} is a fixed point of distributionally robust optimization problems with decision-dependent ambiguity sets, which raises concerns regarding the existence of fixed points and computational tools to compute them. We will resolve them in the remainder of the paper. 

\paragraph*{Uniform robustness levels}
We use the same robustness level $\varepsilon$ and set $\aggregatespace$ for all players. However, all our results readily extend to the case where these are player-specific.

\subsection{Existence of \mbox{Pure Strategically Robust Wardrop Equilibria}}
\label{sec:strategically_robust_wardrop_equilibrium:existence}

In this section, we establish existence of \emph{pure} strategically robust Wardrop equilibria under the \emph{very same} assumptions required for existence of a standard pure Wardrop equilibrium.

\begin{theorem}[Existence]\label{thm:existence}
    Consider a convex aggregative game \game{}, a robustness level $\varepsilon\in\nonnegativereals$, and a compact set $\aggregatespace$.
    Then, a pure strategically robust Wardrop equilibrium exists. 
\end{theorem}

Note that, due to the non-smoothness of the function $x^i\mapsto \sup_{\mu\in\mathcal{B}_\varepsilon(\delta_{\aggregate})}\mathbb{E}^{\aggregatevar \sim \aggregatedistribution} [J^i(x^i, \aggregatevar)]$ for fixed $\sigma\in\reals^n$ (because of the supremum), standard tools from (smooth) variational inequalities~\cite{facchinei2003finite} cannot be applied to establish existence of strategically robust Wardrop equilibria.
In contrast, our proof, deferred to the appendix, follows from recent results in strategically robust game theory, in combination with tools from set-valued analysis and Kakutani's fixed point theorem.

\begin{figure*}[t]
    \centering
    \begin{minipage}[t]{0.495\textwidth}
        \centering
        \begin{minipage}[c][3.8cm][c]{\linewidth}
            \centering






\begin{tikzpicture}
  \begin{axis}[
      width=0.9\textwidth, 
      height=4.0cm,
      axis lines=left,
      axis line style={-stealth, thick},
      enlarge x limits=0.02,
      ymin=5.8, ymax=9.5,
      every major grid/.style={draw=gray!40},
      xlabel={Hour of the day [h]},
      ylabel={Norm. demand [kW]},
      ylabel near ticks,
      xtick={1,3,5,7,9,11,13,15,17,19,21,23},
      xticklabels={12,14,16,18,20,22,0,2,4,6,8,10},
      legend style={
          at={(0.65,0.8)},
          anchor=south,
          legend columns=2,
          font=\scriptsize,
          inner sep=1pt,
          /tikz/every even column/.append style={column sep=0.4cm}
      }
  ]

    \pgfplotstableread[col sep=comma]{results/aggregative_charging_game_N=100_valley_filling.csv}\AggData

    \addplot+[
      ultra thick,
      solid,
      mark=none,
      color=ETHbronze
    ]
    table[
      x=hour,
      y={0.0}
    ]{\AggData};
    \addlegendentry{$\varepsilon=0$};

    \addplot+[
      ultra thick,
      dashed,
      mark=none,
      color=ETHpurple
    ]
    table[
      x=hour,
      y={2.0}
    ]{\AggData};
    \addlegendentry{$\varepsilon=2$};

    \addplot+[
      ultra thick,
      dotted,
      mark=none,
      color=ETHblue
    ]
    table[
      x=hour,
      y={4.0}
    ]{\AggData};
    \addlegendentry{$\varepsilon=4$};

    \addplot+[
      mark=none,
      color=black
    ]
    table[
      x=hour,
      y=non_pev
    ]{\AggData};
    \addlegendentry{non-EV demand};

  \end{axis}
\end{tikzpicture}
        \end{minipage}
        \captionsetup{width=\linewidth}
        \caption[Robust Charging Profiles]{
        Total demand of electricity (i.e., aggregative demand and non-EV demand) for different robustness levels $\varepsilon$.
        Strategic robustness smooths the expected ``valley filling effect'' ($\varepsilon=0$), which eventually disappears when players are overly conservative ($\varepsilon=4$).
        }
        \label{fig:charging_game}
    \end{minipage}\hfill
    \begin{minipage}[t]{0.495\textwidth}
        \centering
        \begin{minipage}[c][3.8cm][c]{\linewidth}
            \centering
            \begin{tikzpicture}
\begin{axis}[
    width=0.95\textwidth,   
    height=4.0cm,
    ybar,
    bar width=6pt,
    grid=major,
    ymin=0,
    enlarge x limits=0.02,
    every major grid/.style={draw=gray!40},
    xlabel={Hour of the day [h]},
    ylabel={Avg. aggregate [kW]},
    xtick={0,2,4,6,8,10,12,14,16,18,20,22},
    xticklabels={12,14,16,18,20,22,0,2,4,6,8,10},
    ylabel near ticks,
    legend pos=north west,
]
\pgfplotstableread[col sep=comma]{results/aggregative_charging_game_N=100_aggregates_by_hour.csv}\AggData

\addplot+[
  bar shift=0pt,
  fill=ETHbronze,   fill opacity=0.5,
  draw=ETHbronze!80!black, draw opacity=0.5
] 
table[
  x expr=\coordindex,   
  y=0.0               
]{\AggData};
\addlegendentry{$\varepsilon=0$};

\addplot+[
  bar shift=0pt,
  fill=ETHpurple,  fill opacity=0.5,
  draw=ETHpurple!80!black, draw opacity=0.5
]
table[
  x expr=\coordindex,
  y=2.0
]{\AggData};
\addlegendentry{$\varepsilon=2$};

\addplot+[
  bar shift=0pt,
  fill=ETHblue,fill opacity=0.5,
  draw=ETHblue!80!black, draw opacity=0.5
]
table[
  x expr=\coordindex,
  y=4.0
]{\AggData};
\addlegendentry{$\varepsilon=4$};

\end{axis}
\end{tikzpicture}
        \end{minipage}
        \captionsetup{width=\linewidth}
        \caption[Robust Charging Profiles]{
        Charging profile for three robustness levels $\varepsilon$ (including the Wardrop equilibrium at $\varepsilon=0$). 
        Strategic robustness flattens the demand for electricity: When players seek protection, they avoid peaks in their energy consumption, as these expose them to larger costs in the presence of perturbed behaviors.
        }
        \label{fig:charging_game_profiles}
    \end{minipage}
\end{figure*}

\subsection{Equivalent Game Formulation}\label{sec:strategically_robust_wardrop_equilibrium:equivalent_game}

Unfortunately, the existence of pure strategically robust equilibria does not suffice to yield a finite-dimensional problem: The equilibrium condition~\eqref{eq:defintion_sr_2} still requires the solution of a maximization problem in the space of probability distributions on $\aggregatespace$, which is infinite-dimensional. 
To bypass this complexity, we follow~\cite[Section 4]{lanzetti2025strategically} and use duality for distributionally robust optimization~(e.g., see~\cite[Theorem 1]{gao2023distributionally}) to obtain a finite-dimensional counterpart of \eqref{eq:defintion_sr_2}.
In particular, we reformulate the problem of finding strategically robust Wardrop equilibria as the problem of finding standard Wardrop equilibria of an appropriately augmented game. 

\begin{proposition}[Equivalent convex aggregative game]\label{prop:equivalent_game_general}
Consider a convex aggregative game \game{}, a compact set $\aggregatespace$, and a robustness level 
$\varepsilon \in \positivereals$.
Let $\mathcal G_\varepsilon$ be the associated convex aggregative game defined by the augmented action spaces 
$
\tilde{\mathcal{X}}_\varepsilon^i \coloneqq \mathcal{X}^i \times [0, M^i],
\quad\text{with}\quad 
M^i\coloneqq 2\max_{x^i \in \mathcal{X}^i,\aggregate\in\aggregatespace} \abs{J^i\bigl(x^i,\aggregate\bigr)}/\varepsilon^\pnorm<+\infty,
$
the same aggregative function \revision{(i.e., $\tilde\sigma_\varepsilon((x^1,\lambda^1),\ldots,(x^N,\lambda^N))=\sigma(x^1,\ldots,x^N)$)},
and the cost functions
\begin{equation}\label{eq:equivalent_game_general}
    \tilde{J}_\varepsilon^i((x^i,\lambda^i),\aggregate)
    \coloneqq 
    \max_{\hat \aggregate \in \aggregatespace}\Bigl\{
    J^i\bigl(x^i,\,\hat \aggregate\bigr)
    -
    \lambda^i \norm{\aggregate-\hat \aggregate}^\pnorm
    \Bigr\}
    +
    \lambda^i\varepsilon^\pnorm.
\end{equation}
Then, $(\wardropeq^1,\dots,\wardropeq^N)$ is a pure 
strategically robust Wardrop equilibrium of \game{} with set $\aggregatespace$ and robustness level~$\varepsilon$ if and only if there exists 
$(\lambda^1,\dots,\lambda^N)$,
with $\lambda^i \in [0,M^i]$ for all $i$, so that 
$\bigl((x_W^1,\lambda^1),\dots,(x_W^N,\lambda^N)\bigr)$ is a pure Wardrop equilibrium of $\mathcal{G}_\varepsilon$.
\end{proposition}

\cref{prop:equivalent_game_general}, which we prove in the appendix, suggests that strategically robust Wardrop equilibria of a game $\mathcal{G}$ can be computed as \emph{standard} Wardrop equilibria of an appropriately modified convex aggregative game $\mathcal{G}_\varepsilon$.
Although the prospect of designing computational tools is enticing, the mere evaluation of the costs $\tilde J^i_\varepsilon$ involves solving a possibly non-convex optimization problem, which is generally prohibitive. 
This complexity prompts us to focus on an instance where we can efficiently handle the maximization problem -- namely when the function $\aggregate\mapsto J^i(x^i,\aggregate)$ is concave for all $x^i\in\mathcal{X}^i$ and $\aggregatespace=\{\aggregate\in\reals^n: f_k(\aggregate)\le0 \,\forall\,k\in\{1,\ldots,K\}\}$ for lower semi-continuous convex $f_k:\reals^n\to\reals\cup\{+\infty\}$.
This setting is motivated by games in markets with linear pricing functions~\cite{ma2011decentralized} or power minimization games in multi-user communication settings~\cite{scutari2010convex}, where the cost is a concave function of the aggregate.
When $J^i$ is concave in $\aggregate$ and $\aggregatespace$ is convex, the maximization problem in~\eqref{eq:equivalent_game_general} is a convex problem. 
We can therefore utilize duality for convex optimization~\cite[Theorem 2]{zhen2023unified} to reformulate the maximization problem in $\tilde J^i_\varepsilon$.

\begin{corollary}[$J^i$ concave in $\aggregate$, $\mathcal{X}_\sigma$ convex]\label{cor:equivalent_game1:computational}
Consider a convex aggregative game \game{}, a robustness level $\varepsilon\in\positivereals$, and a set $\mathcal{X}_\sigma$.
Assume that (i) $\aggregate\mapsto J^i(x^i,\aggregate)$ is concave for all $x^i\in\mathcal{X}^i$ and (ii) $\aggregatespace=\{\aggregate\in\reals^n: f_k(\aggregate)\le0 \,\;\forall\,k\in\{1,\ldots,K\}\}$ for lower semi-continuous convex $f_k:\reals^n\to\reals\cup\{+\infty\}$.
%
Then, $\tilde{J}_\varepsilon^i$ in~\eqref{eq:equivalent_game_general} can be cast as the convex program:
\begin{equation}\label{eq:equivalent_game1:computational}
    \begin{aligned}
    \hspace{-0.5cm}
    \min_{\substack{v,z_k \in \reals^n, \\ \tau_k \in \nonnegativereals}}
    & \lambda^i \varepsilon^\pnorm + (-J^i)^{*,2}(x^i,v) + \sum_{k=1}^K \tau_k f_k^\ast\!\left(\frac{z_k}{\tau_k}\right)
    \\
    &-\aggregate^\top \left(v + \sum_{k=1}^K z_k\right)
    +\lambda^i\psi_p\left(\frac{v + \sum_{k=1}^K z_k}{\lambda^i}\right),
    \end{aligned}
\end{equation}
where the functions $f_k^*(v)\coloneqq \sup_{\aggregate \in \mathbb{R}^n} \left\{ v^\top \aggregate - f_k(\aggregate) \right\}$ and $(-J^i)^{*,2}(x^i,v)\coloneqq\sup_{\aggregate\in\reals^n}\{v^\top\aggregate+J^i(x^i,\aggregate)\}$ are the convex conjugate of $f_k$ and of $-J^i$ (w.r.t. the second argument),  and 
$\psi_p(w)=\frac{(q-1)^{q-1}}{q^q}\norm{w}_*^q$ if $p>1$ 
($\norm{\cdot}_*$ is the dual norm of $\norm{\cdot}$ and $q$ is the conjugate exponent of $\pnorm$, i.e., $\frac{1}{p}+\frac{1}{q}=1$)
and $\psi_{p}(w)=0$ if $\norm{w}_*\leq 1$ and $+\infty$ otherwise if $p=1$. 
\end{corollary}

This reformulation as a minimization problem is attractive since it allows us to augment the players' action space with the variables $\tau_k, v, z_k$ and obtain a game whose costs do \emph{not} ``include'' an optimization problem. As we discuss in~\cref{sec:strategically_robust_wardrop_equilibrium:solution} below, this enables the design of algorithms to compute strategically robust Wardrop equilibria. 
Inspired by our application in EV charging (see~\cref{sec:ev_charging}), \revision{we study aggregative games with affine price functions as an example.} 

\begin{example}[Games with affine prices]\label{ex:affine}
\revision{Let
$
J^i(x^i,\aggregate) \coloneqq \sum_{k=1}^n x_k^ip_k(\aggregatekth),
$
where the price function $p_k$ is $p_k(\aggregatekth) = \alpha_k \aggregatekth+ \beta_k$.}
Let
$\aggregatespace=\{\aggregate\in\reals^n: \aggregatekth\in[0,\revision{\sigma_{\text{max}}}]\}$, $p=2$, and $\norm{\cdot}$ be the Euclidean norm.
Then, the cost function $\tilde J_\varepsilon^i$ reads
\begin{align*}
    \min_{\tau^\mathrm{l},\tau^\mathrm{u} \in \nonnegativereals^n}
    \lambda^i \varepsilon^2 +
    &\sum_{k=1}^n x^i_k\large( \alpha_k\aggregatekth +\beta_k \large) + \aggregatekth \large(\tau^\mathrm{l}_k - \tau_k^\mathrm{u}) + \revision{\sigma_{\text{max}}}\tau^\mathrm{u}_k \notag
    \\
    &+\frac{1}{4\lambda^i}\norm{\tau^\mathrm{l}-\tau^\mathrm{u}+\alpha \odot x^i}^2,
\end{align*}
where $\odot$ denotes element-wise multiplication. We defer the derivation to the appendix. 

\end{example} 

\subsection{Solution Methods}\label{sec:strategically_robust_wardrop_equilibrium:solution}

Armed with an equivalent game reformulation, we now design an iterative algorithm to compute strategically robust Wardrop equilibria. 
We employ an iterative scheme based on \emph{proximal-best responses}, whereby at each iteration players perform a proximal step to minimize their own cost. 
Concretely, at each step \(t\), each player \(i\) selects $(x^i_{t+1}, \lambda^i_{t+1})$ as 
\begin{equation*}
   \argmin_{\substack{
      x^i \in \mathcal{X}^i,\\
      \revision{\lambda^i \in [0,M^i]}
   }}
   \Bigl\{
      \tilde J_\varepsilon^i\bigl((x^i,\lambda^i),\aggregate(x_t)\bigr)
      + \frac{1}{2\rho}\norm{(x^i,\lambda^i) - (x^i_t,\lambda^i_t)}^2
   \Bigr\},
\end{equation*}
where $\rho>0$ can be interpreted as the step size of the scheme. 
Our convex reformulations allow us to write $\tilde J_\varepsilon^i$ as a minimization problem, so a proximal step effectively consists of solving a finite-dimensional optimization problem, which can be done efficiently with off-the-shelf solvers.
If this algorithm converges, then the solution is a strategically robust Wardrop equilibrium. 
\revision{The study of its convergence properties, and, more broadly, decentralized algorithms for strategically robust Wardrop equilibria, is left for future research.}

\begin{figure*}[t]
    \centering
    \vspace{0.6em}
    \begin{minipage}[t]{0.37\textwidth}
         \centering
         \begin{minipage}[c][3.4cm][c]{\linewidth}
             \centering
             \usepgfplotslibrary{groupplots,fillbetween} 
\pgfplotstableread[col sep=comma]{results/aggregative_game_worst_case.csv}\chargingsimdata

\begin{tikzpicture}
  \begin{groupplot}[
      group style={
        group size=3 by 1,
        horizontal sep=1em,
      },
      width=0.49\columnwidth,
      height=4cm,
      xlabel={\(\varepsilon\)},
      xlabel near ticks,
      grid=both,
      xmin=0, xmax=10,
      ymin=66, ymax=80,
      title style={at={(0.5,0.9)}, anchor=south, font=\scriptsize},
  ]
    \nextgroupplot[
      title={1\,\si{\kilo\watt}},
      ylabel={Individual cost [\$]},
      ylabel near ticks,
      legend style={
         font=\scriptsize,
         inner sep=1pt,
         legend columns=1, 
         column sep=1ex
      }
    ]
      \addplot [name path=upper1, ultra thick, color=ETHpurple, mark=none]
        table [x=epsilon, y=worst_individual_cost_1kW] {\chargingsimdata};
      \addplot [name path=lower1, thin, color=ETHpurple, mark=none]
        table [x=epsilon, y=min_individual_cost_1kW] {\chargingsimdata};
      \addplot [fill=ETHpurple!20, forget plot] fill between[of=upper1 and lower1];
      \addplot [ultra thick, color=ETHpurple, mark=none, dotted]
        table [x=epsilon, y=average_individual_cost_1kW] {\chargingsimdata};

      \addlegendimage{ultra thick, mark=none, solid, color=ETHpurple}
      \addlegendentry{Max.}
      \addlegendimage{thin, mark=none, color=ETHpurple}
      \addlegendentry{Min.}
      \addlegendimage{ultra thick, mark=none, dotted, color=ETHpurple}
      \addlegendentry{Avg.}

    \nextgroupplot[
      title={2\,\si{\kilo\watt}},
      ylabel={},
      yticklabels={}
    ]
      \addplot [name path=upper2, ultra thick, color=ETHpurple, mark=none]
        table [x=epsilon, y=worst_individual_cost_2kW] {\chargingsimdata};
      \addplot [name path=lower2, thin, color=ETHpurple, mark=none]
        table [x=epsilon, y=min_individual_cost_2kW] {\chargingsimdata};
      \addplot [fill=ETHpurple!20, forget plot] fill between[of=upper2 and lower2];
      \addplot [ultra thick, color=ETHpurple, mark=none, dotted]
        table [x=epsilon, y=average_individual_cost_2kW] {\chargingsimdata};

    \nextgroupplot[
      title={4\,\si{\kilo\watt}},
      ylabel={},
      yticklabels={}
    ]
      \addplot [name path=upper4, ultra thick, color=ETHpurple, mark=none]
        table [x=epsilon, y=worst_individual_cost_4kW] {\chargingsimdata};
      \addplot [name path=lower4, thin, color=ETHpurple, mark=none]
        table [x=epsilon, y=min_individual_cost_4kW] {\chargingsimdata};
      \addplot [fill=ETHpurple!20, forget plot] fill between[of=upper4 and lower4];
      \addplot [ultra thick, color=ETHpurple, mark=none, dotted]
        table [x=epsilon, y=average_individual_cost_4kW] {\chargingsimdata};

  \end{groupplot}
\end{tikzpicture}
         \end{minipage}
         \captionsetup{width=\linewidth}
         \caption[Charging Worst Case]{
    Average, maximum, and minimum costs when the aggregate demand is increased by 1, 2, and 4\si{\kilo\watt} at random times during the day. }
         \label{fig:charging_worst_case}
    \end{minipage}
    \hfill
    \begin{minipage}[t]{0.3\textwidth}
         \centering
         \begin{minipage}[c][3.4cm][c]{\linewidth}
             \centering

    

\pgfplotsset{compat=1.17}


\begin{tikzpicture}
  \begin{axis}[
    width=1\columnwidth,
    height=4.0cm,
    xlabel={Individual cost [\$]},
    ylabel={Frequency},
    grid=both,
    xmin=66, xmax=74,
    xtick={66,67,68,69,70,71,72,73,74},
    legend pos=north east,
    ybar,
    enlarge x limits=0,
    bar width=1,  
  ]
    \addplot+[
      fill=ETHbronze,
      opacity=0.5,
      draw=black,
      bar shift=0.0
    ]
    table[x=bin_center, y=count_0.0, col sep=comma] {results/histogram_values_uniform.csv};
    \addlegendentry{$\varepsilon=0$}
    
    \addplot+[
      fill=ETHpurple,
      opacity=0.5,
      draw=black,
      bar shift=+0.0
    ]
    table[x=bin_center, y=count_2.0, col sep=comma] {results/histogram_values_uniform.csv};
    \addlegendentry{$\varepsilon=2$}
  \end{axis}

  \path
    (current bounding box.north west) ++(0,3.5mm)
    rectangle
    (current bounding box.south east);
\end{tikzpicture}
         \end{minipage}
         \captionsetup{width=\linewidth}
         \caption[Charging histogram Case]{
        Histogram of individual cost (deviation of 2kW) at the Wardrop equilibrium ($\varepsilon=0$) and strategically robust equilibrium ($\varepsilon=2$).
        }
         \label{fig:charging_cost_histogram}
    \end{minipage}
    \hfill
    \begin{minipage}[t]{0.29\textwidth}
         \centering
         \begin{minipage}[c][3.4cm][c]{\linewidth}
             \centering
             \begin{tikzpicture}
\begin{axis}[
    width=1\columnwidth,
    ylabel near ticks,
    xlabel near ticks,
    height=4cm,
    grid=major,
    every major grid/.style={draw=gray!40},
    xlabel={$\varepsilon$},
    ylabel={Price of anarchy},
    xmin=0, xmax=2, 
    ymin=0.95, ymax=1.40,   
    legend style={
        at={(0.98,0.02)},
        anchor=south east
    }
]

\pgfplotstableread[col sep=comma]{results/aggregative_poa_charging_game_N=100_poa.csv}\PoAdata

\addplot[
    ultra thick,
    no marks 
]
table[
    x=epsilon,
    y=price_of_anarchy
]{\PoAdata};


\end{axis}
\end{tikzpicture}
         \end{minipage}
         \captionsetup{width=\linewidth}
        \caption{Price of anarchy as a function of the robustness level $\varepsilon$. For $\varepsilon=1.2$ the price of anarchy is 1 and players coordinate to minimize social cost.}
         \label{fig:ev:poa:2}
    \end{minipage}
\end{figure*}

\section{EV charging: Coordination via Robustification}\label{sec:ev_charging}

\subsection{Problem Formulation}

\subsubsection*{General problem setting}
We revisit the canonical electric vehicle (EV) charging game studied in~\cite{ma2011decentralized,paccagnan2018efficiency}.
We consider the charging of a population of $N$ plug-in EVs over a horizon of $n$ time periods.
Each EV decides their charging profile $x^i\coloneqq (x^i_1,\ldots,x_n^i) \in \mathbb{R}^n$, where $x^i_k$ denotes the amount charged by EV $i$ at time $k$. 
At each time $k$, $x^i_k$ must be nonnegative and cannot exceed an upper bound $\bar{x}^i_k$, which captures battery constraints and the inability to charge at given times (in which case $\bar x^i_k=0$).
The players aim at charging their batteries to their desired amounts $\theta^i$ by the end of the horizon.
Accordingly, the players' action spaces are
\begin{equation*}
    \mathcal{X}^i
    \coloneqq 
    \left\{ x^i \in \reals^n :
    \begin{array}{l}
    0 \leq x_k^i \leq \bar{x}_k^i, \quad \forall k=1, \ldots, n \\
    \sum_{k=1}^n x_k^i \ge \theta^i
    \end{array}
    \right\}.
\end{equation*}
Each EV seeks to minimize the charging cost, which, at each time, results from the product of the charging $x_k^i$ and the price of electricity at time $k$, denoted by $p_k(\sigma_k)$. We model it as 
$    
    p_k(\sigma_k)
    =
    \frac{\aggregatekth+d_k}{\kappa_k},
$
where $d_k$ represents the non-EV demand, which we take from \cite{ma2011decentralized} and show in~\cref{fig:charging_game}, $\aggregatekth(x)\coloneqq\frac{1}{N}\sum_{i=1}^Nx_k^i$ is the aggregate charging at time $k$, normalized by the number of EVs, and $\kappa_k$ is the normalized total production capacity.
The players' cost functions are then
\begin{equation*}
    J^i(x^i,\aggregate):=\sum_{k=1}^n x_k^ip_k(\aggregatekth).
\end{equation*}
For our strategically robust game, we suppose players do not have full knowledge about the others' action spaces and assume that the normalized aggregate charging lies in 
$
    \mathcal{X}_\aggregate 
    \coloneqq \{\aggregate\in\reals^n: \aggregatekth\in[0,\revision{\sigma_\mathrm{max}}]\}
$
for some $\revision{\sigma_\mathrm{max}}>0$. \revision{Provided it is sufficiently large, it can be shown that the value of $\sigma_\mathrm{max}$ does not affect the strategically robust Wardrop equilibrium.}

\subsubsection*{Our case study}
We consider a time horizon of $n=24$ and a population of $N=100$ homogeneous players. For all players, the upper charging capacity is $\bar x_k^i = 2\si{\kilo\watt}$ between 5pm and 10am and $\bar x_k^i=0$ otherwise (for example because the vehicle is not parked).
The desired charging amount is $\theta^i = 9\si{\kilo\watthour}$; the normalized total production capacity is $\kappa_k=1\si{\kilo\watt}$ for all times.
We compute the strategically robust Wardrop equilibrium using the results in \cref{cor:equivalent_game1:computational,ex:affine} by formulating the proximal optimization problem in CVXPY~\cite{diamond2016cvxpy} and solving it using MOSEK~\cite{mosek-python}\footnote{Code accessible at \url{https://github.com/nicolaslanzetti/strategically-robust-game-theory}.}.

\subsection{Strategically Robust EV Charging Policies}
We show the demand for electricity in~\cref{fig:charging_game_profiles} and the resulting total demand for electricity in~\cref{fig:charging_game} for several robustness levels $\varepsilon$, including the standard Wardrop equilibrium ($\varepsilon=0$). 
For increasing values of $\varepsilon$, players spread their energy consumption throughout the day (see~\cref{fig:charging_game_profiles}). Indeed, during peak hours, when many agents are charging, even slight shifts in the aggregate can result in significantly higher costs for each player. 
For large values of $\varepsilon$ (e.g., $\varepsilon=4$), players almost evenly spread their energy consumption. 
Consequently, as shown in~\cref{fig:charging_game}, the ``valley filling effect'' which arises at the standard Wardrop equilibrium ($\varepsilon=0$) becomes smoother for moderate robustness levels ($\varepsilon=2$).
When players are overly conservative ($\varepsilon=4$), this effect disappears.

To assess the robustness of the various solutions, we consider the following perturbation in the aggregate. At some random time during the day, there is an increase in aggregate consumption $\aggregate$ for two hours.
\revision{This deviation might stem from increased demand of some players, increased base demand, or even a sudden increase in the number of players.} Since the source of uncertainty is generally unknown, we use strategic robustness to directly robustify players against deviations in the emergent aggregate behavior. We conduct this analysis for three magnitudes of the increased consumption (1\si{\kilo\watt}, 2\si{\kilo\watt}, 4\si{\kilo\watt}) and show the resulting cost distributions in~\cref{fig:charging_worst_case,fig:charging_cost_histogram}.
%
\cref{fig:charging_worst_case} shows that some level of robustness leads to an improvement \emph{both} in the average and the worst-case cost, and, more generally, to a more desirable cost distribution, as demonstrated by~\cref{fig:charging_cost_histogram} for the case of 2\si{\kilo\watt}.
This \emph{``coordination-via-robustification''} effect contrasts with the standard intuition in robust optimization: In game-theoretic settings, strategic robustness might yield lower costs for \emph{all} players compared to the standard Wardrop equilibrium.
This effect could be attributed to deviations in the aggregate behavior, as illustrated in~\cref{fig:charging_worst_case}, and is also observed in standard robust optimization. However, notably, it also arises when all the players play their strategically robust strategy without deviating from it, as we investigate in the next section.

\subsection{``Perfect'' Coordination via Robustification}

To conclude, we investigate the role of strategic robustness in one of the settings presented in~\cite{paccagnan2018efficiency}, where there is no base demand (i.e., $d_k=0$) and the price function is constant ($p_k(\aggregatekth) = 0.15$) from 5pm to 1am, and linear ($p_k(\aggregatekth) = 0.15\,\aggregatekth$) from 2am to 10am.
We follow~\cite{paccagnan2018efficiency, ma2011decentralized} and define \emph{social cost} as $\sum_{k=1}^np_k\bigl(\aggregatekth + d_k\bigr)\bigl(\aggregatekth + d_k\bigr)$.
We study the impact of strategic robustness on the price of anarchy, defined as the ratio of the social cost at the strategically robust Wardrop equilibrium to the optimal social cost.
We show the results in~\cref{fig:ev:poa:2}. Remarkably, strategic robustness leads to ``perfect'' coordination of the players for $\varepsilon =1.2$.

\section{Concluding Remarks}

Our application in electric vehicle charging adds to the games studied in~\cite{lanzetti2025strategically} that feature a ``\emph{coordination-via-robustification}'' effect. 
A full characterization of this effect is a natural area for future research. 
To conclude, just as in standard robust optimization, robustness might ease computation, as observed in~\cite{mazumdar2025tractable,zhang2025convergent} in multi-agent reinforcement learning. Accordingly, the study of the computational properties of strategically robust Wardrop equilibria is a promising research direction.


\bibliographystyle{IEEEtran}
\bibliography{references}


\appendix
\subsection{Proof of~\cref{thm:existence}}

We split the proof into four steps. Throughout the proof, let $\mathcal{X}\coloneqq \mathcal{X}^1\times\cdots\times\mathcal{X}^N$ and $\bar{\mathcal{X}}_{\sigma}=\sigma(\mathcal{X})$. In general, $\bar{\mathcal{X}}_\sigma$ need not coincide with $\mathcal{X}_\sigma$, as we defined the former to be the image of $\mathcal{X}$ through $\sigma$ (i.e., $\bar{\mathcal{X}}_\sigma$ is the set of all possible values of $\sigma(x)$) and the latter to be a general compact set (cf.~\cref{assumption_compact}).

\subsubsection*{Step 1: Strategically robust best responses are upper hemicontinuous and have non-empty, compact, and convex values}

For each player $i\in\{1,\ldots,N\}$, we define the strategically robust best response $r_\mathrm{sr}^i:\bar{\aggregatespace} \rightarrow 2^{\mathcal{X}^i}$ as 
\begin{equation*}
    \aggregate\mapsto r^i_\mathrm{sr}(\aggregate)
    \coloneqq 
    \arg \min_{x^i \in \mathcal{X}^i} \bar J^i(x^i,\aggregate),
\end{equation*}
where the function $\bar J^i:\mathcal{X}^i \times \bar{\mathcal{X}}_{\aggregate}\rightarrow \mathbb{R}$ is defined by
\begin{equation*}
    (x^i,\aggregate)
    \mapsto
    \sup_{\mu\in \mathcal{B}_\varepsilon(\delta_{\aggregate})} \mathbb{E}^{\hat\aggregate\sim\mu}[J^i(x^i,\hat\aggregate)],
\end{equation*}
where $\bar{\mathcal{X}_{\aggregate}}$ is a non-empty compact subset of $\reals^n$, being the image of a non-empty compact set through a continuous map.
Moreover, by the proof of~\cite[Theorem 4.2]{lanzetti2025strategically}, the map $\bar J^i$ is continuous in $(x^i,\sigma)$ and convex in $x^i$ for a fixed $\aggregate$ (in particular, the results apply since, by~\cref{assumption_compact}, $\aggregatespace$ is compact).
Then, Berge's Maximum Theorem~\cite[Theorem 9.14]{sundaram1996first} establishes that the best response map $r_\mathrm{sr}^i$ is upper hemicontinuous, nonempty, and compact valued.
Finally, convexity of the set $r_\mathrm{sr}^i(\aggregate)$ for all $\aggregate$ follows from convexity of the function $\bar J^i$ and the action spaces $\mathcal{X}^i$.

\subsubsection*{Step 2: Upper hemicontinuity of $x\mapsto\bar\aggregate(x)\coloneqq\{\aggregate(x)\}$}

By~\cite[Lemma 17.6]{aliprantis2006infinite} and continuity of $\aggregate$, the set-valued map $\bar\aggregate:\mathcal{X}^1\times\cdots\times\mathcal{X}^N\to 2^{\bar{\mathcal{X}}_\sigma}$ which maps each $(x_1,\ldots,x_N)$ to the singleton $\{\aggregate(x)\}\subset\bar{\mathcal{X}}_\sigma$ is upper hemicontinuous.

\subsubsection*{Step 3: The strategically robust best response has closed graph and is non-empty and convex valued}

We define the strategically robust best response map $\Omega: \mathcal{X} \to 2^{\mathcal{X}}$ by 
\begin{equation*}
    \Omega(x)
    \coloneqq 
    \{
        (r^1,\ldots,r^N): r^i\in r^i_\mathrm{sr}(\aggregate(x)) \,\forall\, i\in\{1,\ldots,N\}
    \}.
\end{equation*}
We now prove that $\Omega$ has closed graph. 
The map $\Omega$ can be interpreted as the composition of the set-valued maps $r_\mathrm{sr}^i$ and $\bar\aggregate$, defined in Steps 1 and 2, respectively. 
By Step 1, the correspondence \(r^i_\mathrm{sr}\) is upper hemicontinuous with nonempty and closed (or compact) values for each player \(i\).
By Step 2, $\bar\aggregate: \mathcal{X} \to 2^{\bar{\mathcal{X}}_\sigma}$ is upper hemicontinuous.
Then, we can leverage \cite[Theorem 17.23]{aliprantis2006infinite} to conclude each composition \(r^i_\mathrm{sr} \circ \bar\aggregate\) is upper hemicontinuous.
Since \(\mathcal{X}^i\) is Hausdorff (being a subset of $\reals^n$) and the values of $r^i_\mathrm{sr}$ are closed (by compactness), \cite[Theorem 17.11]{aliprantis2006infinite} implies that the graph of each \(r^i_\mathrm{sr} \circ \bar\aggregate\) is closed.
Consequently, the graph of \(\Omega\), being the Cartesian product of these closed graphs, is itself closed.
To prove that $\Omega$ has non-empty convex values, it suffices to observe that, for each $i$ and $\aggregate$, the set of best responses $r_\mathrm{sr}^i(\aggregate)$ is non-empty and convex. Since the Cartesian product of non-empty convex sets is itself non-empty and convex, we conclude.

\subsubsection*{Step 4: Kakutani's fixed point theorem}

Since \(\Omega: \mathcal{X} \rightarrow 2^\mathcal{X}\) has (i) non-empty compact and convex domain, (ii) nonempty convex values, and (iii) a closed graph, we can use Kakutani's Fixed Point Theorem (e.g., see~\cite[Corollary 17.55]{aliprantis2006infinite}) to ensure the existence of an \(x^* \in \mathcal{X}\) such that \(x^* \in \Omega(x^*)\).
This is a strategically robust Wardrop equilibrium.

\subsection{Proof of~\cref{prop:equivalent_game_general}}\label{prop:equivalent_game_general:proof}

The proof is analogous to the proof of~\cite[Proposition 4.3]{lanzetti2025strategically}.

\subsection{Proof of~\cref{cor:equivalent_game1:computational}}

This dual reformulation follows directly from~\cite[Theorem 2]{zhen2023unified}. For $p>1$, the convex conjugate of a general norm to the power of $\pnorm$ is given as in \cite[Lemma 6.19]{kuhn2024distributionallyrobustoptimization}. For $p=1$,  the convex conjugate of $d(\aggregate,\hat\aggregate)=\norm{\aggregate-\hat\aggregate}$ is $d^{*,2}(w,\aggregate) = w^\top \aggregate$ if $\norm{w}_* \le 1$ and $+\infty$ otherwise~\cite[Example 3.26]{boyd2004convex}.

\subsection{Proof of~\cref{ex:affine}}
We select $p=2$ and adopt the standard Euclidean norm. The Euclidean norm is self-dual (i.e., $\norm{\cdot}_*=\norm{\cdot}$) and $p=q=2$.
To start, $\aggregatespace$ can be expressed via $f_k^\mathrm{l}(\aggregate)=-\aggregate_k$ and $f_k^\mathrm{u}(\aggregate)=\aggregate_k-\revision{\sigma_{\text{max}}}$ (i.e., $K=2n$).
To obtain the result, it suffices to derive the convex conjugates of $f_k^\mathrm{u}$and substitute them into the formulation of \cref{cor:equivalent_game1:computational}; in particular, the definition of convex conjugate directly yields $(-J^i)^{*,2} (x^i,v)=\sum_{k=1}^n \beta_kx_k^i$ if $v + \alpha \odot x^i = 0$ and $+\infty$ otherwise, $(f_k^\mathrm{u})^* (\frac{z_k}{\tau_k^\mathrm{u}})=\revision{\sigma_{\text{max}}}$ if $\frac{z_k}{\tau_k^\mathrm{u}} = e_k$ and $+\infty$ otherwise, and $(f_k^\mathrm{l})^* (\frac{z_k}{\tau_k^\mathrm{l}})=0$ if $\frac{z_k}{\tau_k^\mathrm{l}} = -e_k$ and $+\infty$ otherwise.

\end{document}